\documentclass[12pt]{iopart}\usepackage{iopams}  

\usepackage{graphicx}%
\usepackage{epsfig}
\usepackage{amssymb}
\usepackage{eucal} 

\begin{document}
\title[Inference of Schr\"odinger Equation]{Inference of Schr\"odinger Equation from Classical-Mechanical Solution
}
\author{J.X. Zheng-Johansson$^{1}$ and P-I. Johansson$^{2}$
}

\date{May 18, 2007}

\address{1. IOFPR,   611 93 Nyk\"oping, Sweden }
\address{2. Uppsala University, 611 82 Nyk\"oping, Sweden 
}
\address{Augmented improved May, 2007}


\begin{abstract}
We set up the classical wave equation for a particle formed of an oscillatory zero-rest-mass charge together with its resulting electromagnetic waves, traveling in a potential field $V$ in a susceptible vacuum. The waves are Doppler-displaced upon the source motion, and superpose into a total, traveling- and in turn a standing- beat wave, or de Broglie phase wave, described by a corresponding total classical wave equation. By back-substitution of the explicitly known total, standing beat wave function and upon appropriate reductions at classic-velocity limit, we separate out from the total a component wave equation describing the kinetic motion of particle, which is equivalent to the Schr\"odinger equation.  The Schr\"odinger wave function follows to be the envelope function of the standing beat wave at classic-velocity limit. 

\end{abstract}


\pacs{  
 03.50.De, 
 03.65.Ta, 
 03.00.00, 
 04.20.Cv, 
 04.30.Db, 
11.00.00, 
41.60.-m, 
41.20.Jb 
}

\maketitle 

\def\citeRef1{7\hspace{-0.1cm}}
\def\Unifcite{7\hspace{-0.1cm}}
\def\Omegavel{\mathbin{{\mit\Omega}\mkern-13.mu^{_{\mbox{$-$}}}\hspace{-0.08cm}{}_d }}

\def\App{}
\def\Thm{{\mit{\Theta}}}

\def\Xim{\chi}

\def\Mcal{{\mathfrak{M}}}
\def\vel{\upsilon}
\def\lb{{\bf l}}
\def\vb{{\bf v}}

\def\Rb{{\bf R}}
\def\pd{\partial}
\def\vphi{\varphi}
\def\psiR{\widetilde{\psi}}
\def\psiL{\widetilde{\psi}^{{\rm vir}}}
\def\PhimR{\widetilde{ {\mit \Phi}}}
\def\PsimR{\widetilde{ {\mit \Psi}}}
\def\PsimL{{\widetilde{ {\mit \Psi}}}^{{\rm vir}}}
\def\a{\alpha}
\def\uav{\bar{u}}
\def\D{\Delta}
\def\th{\theta}
\def\r{{\mbox{\tiny${R}$}}}
\def\re{{\mbox{\tiny${R}$}}}
\def\Fmed{F_{{\rm a.med}}}
\def\med{{\rm med}}
\def\Lw{L_{\varphi}}

\def\Efb{{\bf E}}
\def\Bfb{{\bf B}}
\def\Ac{ \varphi}
\def\Xsub{{\mbox{\tiny${X}$}}}

\def\Ksub{{\mbox{\tiny${K}$}}}
\def\W{{\mit \Omega}}
\def\Wd{\W_d{}}
\def\Nu{{\cal V}}
\def\Nud{\Nu_d{}}
\def\Eng{{\cal E}}
\def\eng{{\varepsilon}}
\def\Acuni{\Ac_{{\Ksub}^\dagsup}^{\dagsup}}
\def\unduni{\Ac_{{\Ksub}^\dagger}^{\dagsup}}
\def\Acauni{\Ac_{{\Ksub}^\ddagsup}^{\ddagsup}}
\def\Acunim{{\Ac_{{\Ksub}^\dagsup}^{\dagsup *}}}
\def\undunim{{\Ac_{{\Ksub}^\dagsup}^{\dagsup}}^*}
\def\Acaunim{{\Ac_{{\Ksub}^\ddagsup}^{\ddagsup *}}}
\def\pd{\partial}
\def\Ad{ {\mit \psi}}
\def\psim{ {\mit \psi}}
\def\Kd{K_d{}}
\def\Lam{{\mit \Lambda}}
\def\lam{\lambda}
\def\dagsup{{\mbox{\tiny${\dagger}$}}}
\def\ddagsup{{\mbox{\tiny${\ddagger}$}}}
\def\psimKdK{\psim_{\Ksub,\Kdsub}}
\def\w{\omega{}}
\def\wdlow{\omega_d }
\def\g{\gamma{}} 
\def\Phim{{\mit \Phi}}
\def\Psim{{\mit \Psi}}
\def\arm{{\rm a}}
\def\brm{{\rm b}}
\def\crm{{\rm c}}
\def\drm{{\rm d}}
\def\erm{{\rm e}}
\def\frm{{\rm f}}
\def\grm{{\rm g}}
\def\hrm{{\rm h}}
\def\lf{\left}
\def\rt{\right}
\def\Kdsub{{\mbox{\tiny${K_d}$}}}
\def\psimkd{\psim_{\kdsub}}
\def\psimKd{\psim_{\Kdsub}}
\def\hquad{ \ \ } 
\def\Taum{{\mit \Gamma}}

\section{Introduction}

Schr\"odinger equation\cite{Schr1926}  in real space, together with the Heisenberg-Born matrix formalism  \cite{Heisenberg:1925} in momentum space, has proven to be the governing law of particle dynamics  in small geometries in nonrelativistic regime. However, up to the present  the nature, 
or 
the interpretation of Schr\"odinger wave function and more generally of quantum mechanics, 
and a corresponding derivation of the  Schr\"odinger equation on the basis, in the first place, of
the elementary classical-mechanical laws for point objects,   have remained as two major unsettled aspects of quantum mechanics.
     E. Schr\"odinger developed his wave mechanics  upon a  vision of the analogy of quantum mechanical waves with classical waves\cite{Schr1926}, though in a phenomenological fashion. So was the earlier inspiring work of L. de Broglie's hypothetical 
matter wave\cite{deBroglie:1924}. 
Since the foundation of quantum mechanics in the 1900s-1930s, various interpretations have been put forward. 
One which  embraces a scheme for effectively deriving the Schr\"odinger equation, to mention only here, is the stochastic electrodynamics approach
\cite{Madelung:1927,Nelson:1966}; this has also served an effective means for deriving the Doebner-Goldin nonlinear Schr\"odinger equation\cite{DoebnerGoldin:1992}.

In view of a scheme where the quantum-mechanical variables and states 
are the result  of the {\it internal processes} of a particle,  
however, 
no success going beyond L. de Broglie's hypothetical phase wave has appeared feasible[\citeRef1 a] prior to our recent 
unification work [\citeRef1 a-k]. 
At the basis of the unification work is a  {\it basic particle formation} (BPF) scheme [\citeRef1 b-e] for basic, or simple (material) particles like the electron, proton, etc. 
 and a model structure of the vacuum [\citeRef1 b,f-g], derived  with the  overall well established experimental observations as input data.
A basic particle according to the BPF scheme, briefly, is formed of an oscillatory massless elementary charge of a specified sign termed a (single) vaculeon,  and the resulting electromagnetic waves in the vacuum.  The vacuum according to the model construction is filled of electrically neutral but polarizable  entities termed vacuuons, each constituted of a p-vaculeon   of charge $+e$ at the core and n-vaculeon of charge  $-e$ on the envelope, bound with each other by an immense  Coulomb energy.

A single vaculeon when disintegrated from a bound state,  vacuuon, is endowed with a certain excess kinetic energy $\eng_q$; this energy can not be dissipated in the vacuum where there is no lower energy levels for it to transit to except in a pair annihilation. In a vacuuonic vacuum where the  vacuuons about the single vaculeon charge are polarized and form a potential well to the latter, $\eng_q$ manifests as a mechanical energy. The vaculeon charge will execute oscillations in the potential well about a fixed site if $\eng_q$ is lesser than the energy barrier with a neighboring site, and also translate if $\eng_q$ greater[\citeRef1\ b]. With a specified $\eng_q$ of a single vaculeon charge as the sole input data,  the first principles classical-mechanical solutions for the resultant material particle  in terms of the internal processes of the particle, have predicted [\citeRef1\ a-k]  a range of observational basic particle properties, including  mass[\citeRef1 b], spin[\citeRef1 b],  the de Broglie wave function[\citeRef1 b-c], the de Broglie relations[\citeRef1 b-c], the Schr\"odinger equation[\citeRef1 a-b,h], the Doebner-Goldin nonlinear Schr\"odinger equation[\citeRef1 i], the Dirac equation[\citeRef1\ j], the Einstein mass-energy relation[\citeRef1 b,d], the Newton's law of gravitation[\citeRef1 h,k], and the Galilean-Lorentz transformation[\citeRef1 h], among others, under corresponding conditions.

In this paper we elucidate in a self-contained fashion in detail the inference of Schr\"odinger equation; this is an augmented treatment of the problem  reported in a symposium paper [\citeRef1 a] earlier, and at greater length in a chapter in book [\citeRef1 h]. The BPF scheme and the FPs classical-mechanics solutions enable a range of the outstanding fundamental problems to be comprehended on a unified theoretical basis. In immediate  relevance to this paper, it is pointed to that it is the vacuum polarized perturbed by the charge of a particle that is waving in the Schr\"odinger wave function of the particle and more simply its de Broglie wave. This  periodic  motion of the vacuum in the original, pre-superposition form corresponds directly to the electromagnetic waves generated by the charge. The mass of a particle is the result of the total motion of the  charge or alternatively the waves of the particle against a  frictional vacuum.

\section{Classical wave equation for component electromagnetic  waves}\label{Sec-QMdr-o-pot}

We consider a particle fabricated according to the BPF scheme [\citeRef1 b, c-g] from a single oscillatory vaculeon charge, $|q|=e$, of zero rest mass, and the resulting electromagnetic waves of frequency $\w^j$. Let the charge and thus the particle be  traveling at a velocity $\vel$, for simplicity firstly in a one-dimensional box of side $L$ along $X$-axis, against an applied conservative potential field $V$. The particle will acquire an inertial mass $m$ dynamically as the result of the total motion of its charge  or alternatively its waves.

We shall here describe the electromagnetic waves in mechanical terms  in the vacuuonic vacuum. Subjecting to the static field of the charge this vacuum is firstly  polarized and induced with a shear elasticity. Subjecting in turn to the disturbance of the charge oscillation,  the now elastic vacuum is consequently produced with a periodic  deformation, of transverse ($Z$-direction) displacements $u^{\dagsup} =a_1 \varphi^{\dagsup} (X,T) $ and $u^{\ddagsup} =a_1 \varphi^{\ddagsup} (X,T)$, generated to  the charge's right and left, i.e. to the $+X$- and $-X$-directions, in the linear chain of vacuum along the box here, $a_1$ being a conversion factor of length unit, and  $\varphi^{\dagsup} $ and $\varphi^{\ddagsup}$ dimensionless. $u^{j} $ corresponds to a "radiation displacement"  out of a total displacement $u_{tot}$ of the vaculeon charge driving by its total energy $\eng_q$, with $u_{tot}>> u^{j}$ in general (see an elucidation of the dynamical process in [\citeRef1 b]). Upon the deformation
the vacuum is accordingly subject to a tensile force $F_\r=c^2 \rho_l$, with $c$ the wave velocity  at which $\varphi^j $ propagates and is equal to the velocity of light for $\varphi^j $ to be identified to the electromagnetic wave, and $\rho_l$ the linear (dynamic) mass density of the vacuum along the box. Besides, $V(X,T)$, and the associated force $F=-\frac{\pd V}{\pd X}$ acting on the charge map to a potential  $V_\med$ and force 
$$\displaylines{
\refstepcounter{equation}\label{eq-Fmeda}
\qquad
F_\med= - \partial V_{\med}/\partial X=-  V_{\med}(V(X,T))/\Lw \hfill (\ref{eq-Fmeda})
}$$ 
which act on the medium. Where the last expression holds because the force acting at one location is uniformly transformed to all other locations in a continuum vacuum here in a chain;  $\Lw$ is the total length of the trains of the resulting waves generated by a continuous span of  oscillation of the charge if without re-absorbing the reflected waves; $\Lw=JL$ if $\Lw$ winds across $L$ in $J$ loops.  $V(X,T)$ in the above ought to write  as $V(X(t'),t')$, its action includes the switching on of the potential from zero at time $t'=0$ to the present value $V$, hence 
$F=-\frac{\partial V(X(t'),t')}{\partial X} = -\frac{\partial V(X,t')}{\partial X}- \frac{\partial V(X(t'),t')}{\partial t'}
\frac{1}{\partial X(t')/ \partial t'} $. 
As a physical requirement the acceleration produced to the mass $m$ of the particle and (effectively) to the mass $\Mcal_\varphi$, $=\Lw \rho_l$,   of the vacuum  must equal, that is
$\partial^2 X/\partial T^2 = -(1/m)(\partial V/\partial X)  $ $ = -(1/ \Mcal_\varphi)(\partial V_{\med}/\partial X)$. Or, $V_{\med}
=(\Mcal_\vphi/m) V 
=\rho_l \Lw V/m 
$. 
With this in (\ref{eq-Fmeda}) we have  
 $$\displaylines{\refstepcounter{equation}\label{eq-Fmed}
\qquad
F_\med 
= -\rho_l V/m.  \hfill (\ref{eq-Fmed})
}$$

The positive of $F_{\r}$ is a tension while the positive of $F_\med$ is a contraction which on the axis of tension is thus negative. The total force acting at a point, uniformly across the chain 
assuming $\varphi^j$ is small, 
is therefore
$$\displaylines{
 \refstepcounter{equation}\label{eq-Fp} 
\qquad 
F'_{\r} = F_{\r}-  F_{\med} = \rho_l c^2  +\rho_lV/m.  
\hfill (\ref{eq-Fp})
}$$  
Consider a segment $\D L$ of the chain
 is tilted from the $X$-axis, in genral differently,  
at an angle  $\theta^j $ and $\theta^j+\D \theta^j $ at its ends $X$  and $ X+\D X$, these being in turn dependent on the wave variables and thus (see below) the directions of travel of the corresponding waves, $j=\dagger$ and  $\ddagger$ as specified earlier. With $\sin \theta^j = \frac{\D (a_1 \varphi^j)}{\D L}$ and $\sin (\theta^j+\D \theta^j )-\sin (\theta^j) \simeq \D \theta^j = \frac{\partial^2 (a_1 \varphi^j)}{\partial X^2} \D X$, where the approximation holds good for $\varphi^j$ small, 
the transverse ($Z$-) component of the net force on $\D L$, $\D F_{\r.t}^{\prime j}= F'_\r [\sin (\theta^j+\D \theta^j )-\sin (\theta^j)]$,  is  accordingly:
$$\displaylines{
\refstepcounter{equation}\label{eq-Frp}
\qquad 
\D F_{\r.t}^{\prime j}
=\left(\rho_l c^2  +\frac{\rho_l V}{m }  \right)\frac{\partial^2 (a_1 \varphi^j) }{\partial X^2} \D X. 
\hfill (\ref{eq-Frp}) 
}  $$
Newton's second law applied to  the segment of mass $\rho_l \D L$, $\simeq \rho_l \D X$, writes $\D {F^j}'_{\r.t} = \frac{ \partial ^2 \varphi^j}{\partial T^2} \rho_l \D X $. With (\ref{eq-Frp}) for $\D {F^j}'_{\r.t} $, dividing $\rho_l \D X$ through, we then obtain the  equation of motion for per unit length per unit density  of the chain, or alternatively the classical wave equation for a component total wave $j$ of the particle:
$$\displaylines{
 \refstepcounter{equation}\label{eq-CME1}
\qquad 
c^2 \frac{\partial^2  \varphi^j(X,T)}{ \partial X^2}  + \frac{ V(X,T)}{m}\frac{\partial^2  \varphi^j(X,T)}{ \partial X^2} =\frac{ \partial ^2 \varphi^j(X,T)}{\partial T^2}, \qquad j=\dagger, \ddagger  
                    \hfill (\ref{eq-CME1})
}$$
$\varphi^j(X,T)$ follows from our representation to be a mechanical description of the electromagnetic wave with  a radiation electric field $E^j$ and magnetic field $B^j= E^j/c$. These are related as $a\varphi^j(X,T) =f E^j(X,T)$ through a conversion factor $f$ (see [\citeRef1 h] for an explicit expression). Accordingly (\ref{eq-CME1}) substituted with $E^j$ for $\varphi^j$ gives the classical wave equation of the corresponding electromagnetic wave $j$ (usually with $V=0$) directly derivable from the Maxwell's equations.

\section{Component total wave solution. Particle total wave from explicit superposition}\label{Sec-V0.1}

\subparagraph{\ref{Sec-V0.1}.1 
Wave displacements; wave parameters; Doppler effect due to source motion} 

\begin{figure}[hb]
\vspace{0.cm}
\begin{flushright}
\includegraphics[width=0.9\textwidth]{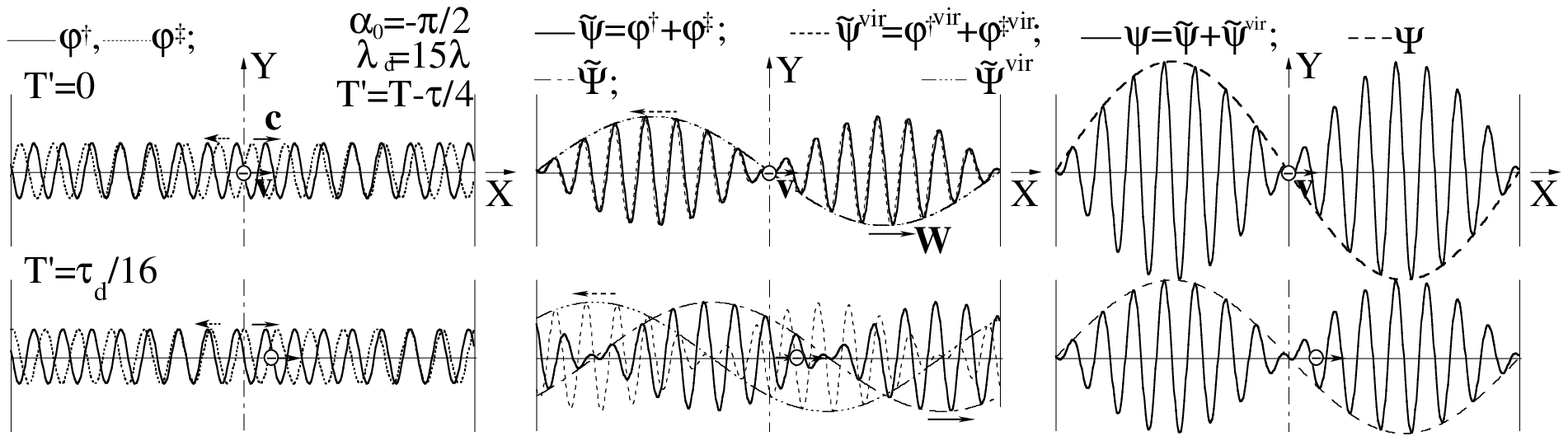}
\end{flushright}
\vspace{-1.05cm}
\begin{flushright}
\includegraphics[width=0.905\textwidth]{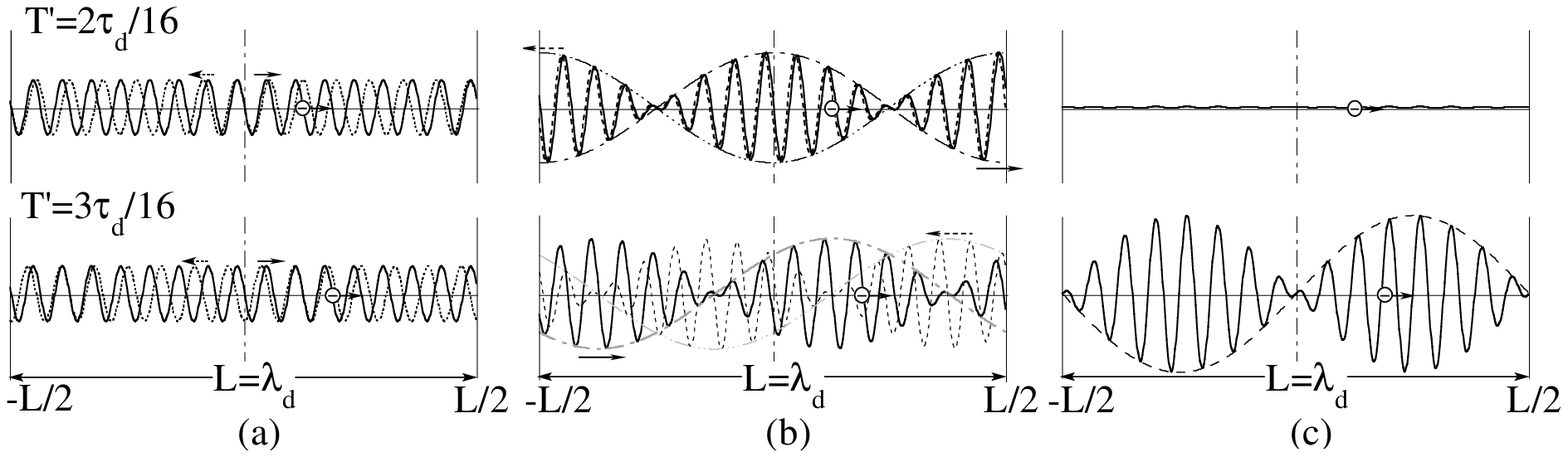}
\end{flushright}
\vspace{-0.5cm}
 \caption{   (a) shows the time development in $3\tau_d/16$,  of a basic particle formed of a traveling oscillatory charge $\ominus$  and the resulting electromagnetic (or alternatively mechanical) waves $\vphi^j$ with wavevectors $k^j$ Doppler displaced due to the source motion at the velocity $\vel$.
 In (b), the opposite traveling $\vphi^{\dagsup}$ and $\vphi^{\ddagsup}$ due to the actual charge, and  ${\vphi^{\dagsup}}^{ {\rm vir}}$ and ${\vphi^{\ddagsup}}^{ {\rm vir}}$ the reflected charge (not shown)
 superpose respectively to 
 the beat waves, or alternatively de Broglie phase waves $\psiR $ and $\psiL$ traveling to the right and left at the phase speed $W $. In (c), $\psiR $ and $\psiL$  superpose  to a standing beat or de Broglie phase wave $\psi$. The envelope functions 
${\widetilde \Psim}$, ${\widetilde \Psim}{}^{{\rm vir}}$ and  $\Psim$,  of a wavelength $\lam_d$ each, viewed together with the source's $\vel$- motion, describe the traveling and standing de Broglie waves of the particle.   
}  
\label{fig-dBwav}
\end{figure}

Let firstly  $V(X,T)$  equal (or approximately equal) a constant, $V_0$.  So the solutions for (\ref{eq-CME1}) are (to a good approximation) plane waves generated in the $+X$- and $-X$- directions (Figure \ref{fig-dBwav}a):   
$$\displaylines{\refstepcounter{equation}\label{eq-Eq3}
\qquad
\varphi^{\dagsup}= C_1 \sin(k^{\dagsup} X- \w ^{\dagsup} T + \a_0), \quad 
\varphi^{\ddagsup}
= -C_1 \sin(k^{\ddagsup} X+ \w ^{\ddagsup} T - \a_0 ), \hfill (\ref{eq-Eq3})
}$$
with $\a_0$ the initial  phase. $\varphi^{\dagsup}$  is  surpassed by its traveling source charge by a distance $\vel \Taum $, and  $\varphi^{\ddagsup}$ is receded by the charge by  $-\vel \Taum$,  in each period $\Taum$. Therefore, their respective wavelengths are Doppler-displaced  to $\lam^{\dagsup}=(1-\vel/c)\Lam $ and  $\lam^{\ddagsup}=(1+\vel/c)\Lam $ 
from $\Lam$ ($=\Taum c$) due to the same source but with $\vel=0$,
as measured by an observer at rest in the vacuum. Accordingly their wavevectors and angular frequencies are displaced to
$$\displaylines{ \refstepcounter{equation}\label{eq-wpara1}
\qquad
k^{\dagsup}=\g^{\dagsup} K
=K+k_d^{\dagsup},  
 \  
k^{\ddagsup}=\g^{\ddagsup}K
=K-k_d^{\ddagsup};
\quad
\w^{\dagsup}=\g^{\dagsup} \W, 
 \  
\w^{\ddagsup}=\g^{\ddagsup}\W; \hfill (\ref{eq-wpara1})
}$$
where  
$\g^{\dagsup}=1/(1-\vel/c), 
\g^{\ddagsup}=1/(1+\vel/c)$;
$K=\frac{2\pi}{\Lam}$, $\W=2\pi/\Taum=cK$; 
 and 
$$\displaylines{
\qquad
k_d^{\dagsup}=\g^{\dagsup} K_d, \
\   k_d^{\ddagsup}=\g^{\ddagsup} K_d;
\quad
\refstepcounter{equation}\label{eq-Eq5}
  K_d =\lf(\frac{\vel}{c}\rt)K. 
                  \hfill  (\ref{eq-Eq5})
}$$

\subparagraph{\ref{Sec-V0.1}.2 Dynamical variables of particle}

The dynamical variables of the total wave, and therefore of the particle  are appropriately the geometric means of the respective  conjugate Doppler-displaced wave variables that represent two wave processes mutually stochastic and sampled at one time. We will involve such  variables relating to the total motion of the particle as given in (\ref{eq-w1}), and to its kinetic motion as in (\ref{eq-Eq5a}) below:
$$\displaylines{
\refstepcounter{equation}\label{eq-w1}
\qquad 
\lam=\sqrt{\lam^{\dagsup}\lam^{\ddagsup}}= \Lam/\g,  
\quad
k=\sqrt{k^{\dagsup}k^{\ddagsup}}=\g K, 
\quad
 \w=\sqrt{\w^{\dagsup}\w^{\ddagsup}}= \g \W; \hfill (\ref{eq-w1})
\cr
\refstepcounter{equation}\label{eq-Eq5a}
\label{eq-Eq6}
 \qquad 
k_d=\sqrt{(k^{\dagsup}-K)(K-k^{\ddagsup})}
     =\g K_d, \hfill
\cr
\qquad
\w_d = k_d \vel 
=\g \W_d;
\quad
\W_d
          = K_d \vel
=\left(\frac{\vel}{c}\right)^2 \W. 
                     \hfill (\ref{eq-Eq6})
}$$
Where $\g=\sqrt{\g^{\dagsup}\g^{\ddagsup}}=1/\sqrt{1-\vel^2/c^2}$; for obtaining the final expressions of the relations of (\ref{eq-w1})--(\ref{eq-Eq6}), the relations of (\ref{eq-wpara1})--(\ref{eq-Eq5}) are used. The energies of the $\vphi^{\dagsup} $ and $\vphi^{\ddagsup} $ waves  are according to  M. Planck $\hbar \w^{\dagsup}$ and $\hbar \w^{\ddagsup}$. It follows from the Planck energies and (\ref{eq-w1}), that $\hbar \w=\hbar \sqrt{\w^{\dagsup} \w^{\ddagsup}}$ is therefore the total energy of the total wave and accordingly the particle. The total energy of the particle $\hbar \w$ is in turn related to its  relativistic mass $m$ according to A.Einstein's mass-energy relation $\hbar \w =m c^2$, ---given also as direct classical-mechanical solution in [\citeRef1 c,d] in terms of the BPF scheme. Thus
  $$\displaylines{
 \refstepcounter{equation}\label{eq-mM}
\qquad m=\hbar \w /c^2, \quad 
 m=\hbar(\g \W)/c^2=\g M; \quad {\rm or} \quad c^2 =\frac{\hbar \w }{m}. 
  \hfill (\ref{eq-mM})}$$

\subparagraph{\ref{Sec-V0.1}.3 Particle total wave from explicit superposition}\label{Sec-V0.2}
$\varphi^{\dagsup}(X,T)$ generated at location $X$ 
and time $T$ to the right of the charge will after a round trip between the infinite walls of the one-dimensional box of a distance $2L$ return to the charge from  left at $X$ at $T+\D T$
(for $\vel<< c$, the source is basically not moved in time $\D T$). Here it meets $ \varphi^{\ddagsup}(X,T+\D T)$ just generated to the left, and superposes with it as $\psiR= \varphi^{\dagsup}(X+2L,T+\D T)+ \varphi^{\ddagsup}(X,T+\D T)$. 
Because $ \varphi^{\dagsup}$  has travelled 
an extra distance $2L$, though having the same time lag $\D T$ as $ \varphi^{\ddagsup}$, the two waves acquire an average relative phase $\a'\simeq 2L K$. If $2L K= N2\pi$,  $N=$ integer, then $\psiR$ is a maximum.
 With some algebra using the trigonometric identity, reorganizing, the total wave displacement writes 
 $$\displaylines{
\refstepcounter{equation}\label{eq-psiRaa}
\qquad
\psiR
=2C_1 \cos[KX+\frac{1}{2}(k_d^{\dagsup}-k_d^{\ddagsup})X-\frac{1}{2}(\w^{\dagsup}-\w^{\ddagsup})T]
\sin\frac{1}{2}[(k_d^{\dagsup}+k_d^{\ddagsup})X-(\w^{\dagsup}+\w^{\ddagsup})T]. 
\hfill (\ref{eq-psiRaa})
}$$
Where, 
$
 k_d^{\dagsup}-k_d^{\ddagsup} 
=\g 2\lf(\frac{\vel}{c}\rt)k_d \dot{=}2\lf(\frac{\vel}{c}\rt)k_d 
$, \ 
$\w^{\dagsup}-\w^{\ddagsup}
=\g 2\lf(\frac{\vel}{c}\rt)\w \dot{=}2\lf(\frac{\vel}{c}\rt)\w
$,  \ 
$
 k_d^{\dagsup}+k_d^{\ddagsup}=\g 2k_d\dot{=}2k_d$, 
\ 
$\w^{\dagsup}+\w^{\ddagsup}=\g 2\w\dot{=}2\w   
$. Of the third and the fourth relations, the $\g$ factor in front of $k_d$ and $\w$ 
represents a higher-order relativistic effect  and has been dropped in the final results, valid for the classic-velocity limit in question here. 
The two leading-term final results, $k_d=\g K_d$ and $\w=\g K$  represent 
each significant relativistic dynamical variables  of the particle; the $\g$ factor in them contains a portion of the kinetic motion  of the particle which is a {\it classical} quantity (see further the discussion after equation (\ref{eq-CME1p2}) later  and in \ref{App-A}) and will  for now be  retained in full. Of the first two relations, the resulting quantities $\lf(\frac{\vel}{c}\rt)k_d $ and $\lf(\frac{\vel}{c}\rt)\w$ are physically
meaningful  only as the order $\vel/c$-quantities  of $k_d$ and $\w$; the factor $\g$ 
further to these makes these even higher order  in $\vel/c$, and has therefore been dropped in the final results.   

Substituting now with the above four relations in (\ref{eq-psiRaa}), 
$\psiR$ writes as  
 $$\displaylines{ \refstepcounter{equation}\label{eq-psiRa}
\hfill \psiR=
{\widetilde\Phim}{\widetilde\Psim}, \quad
{\widetilde\Phim}=2C_1\cos[(K + \lf(\frac{\vel}{c}\rt) k_d)  X  - \lf(\frac{\vel}{c}\rt) \w  T],
\quad
{\widetilde\Psim}=
\sin[ k_d X - \w T+\a_0]. 
               \hfill
(\ref{eq-psiRa})
}$$
The time development of $ \psiR$ is graphically shown by the solid curves in 
Figure \ref{fig-dBwav}b  in a duration $(3/16)\tau_d$, with $\tau_d=2\pi/\w_d$. 
We see that,  $\psiR$  is a product of two parts.  Of them   
${\widetilde\Phim}$  rapidly oscillates in space, to a leading term $\cos (KX)$, at the wavelength $\Lam=2\pi/K$.  This is modified by a higher-order term 
  $ (\vel/c) k_d  X  - (\vel/c)  \w  T$, so that  
${\widetilde\Phim}$ presents a slow phase velocity $W_2=\frac{(\vel/c)\g \w}{K+(\vel/c)\g k_d} \simeq \frac{(\vel/c) \W}{K}=\vel$, reflecting the source translation. 
 The second part, ${\widetilde\Psim}$ (single-dot-dashed curve in Figure \ref{fig-dBwav}b) envelops the rapid oscillating ${\widetilde\Phim}$, and oscillates in space at the wavelength $\lam_d=2\pi/k_d$.

Overall, with ${\widetilde\Phim}\simeq\cos (KX)$,  $\psiR$ describes a traveling beat wave,  
of a wavevector $k_d$, 
wavelength  $\lam_d=2\pi/k_d$, angular frequency  $\w$, 
and  phase velocity 
$$\displaylines{
 \refstepcounter{equation}\label{eq-W}
  \qquad               W= \w/k_d=c^2/\vel.  
             \hfill (\ref{eq-W})
}$$  
Accordingly ${\widetilde\Psim}$ represents the envelope function of the traveling beat; effectively,
${\widetilde\Psim}$  
describes the total wave $\psiR$ if ignoring the detailed oscillation, ${\widetilde\Phim}$, in $\psiR$. 
In view that ${\widetilde \Psim}$, or $\psiR$ in terms of its ${\widetilde \Psim}$, taken together with the source motion of velocity $\vel$ 
 describes the  kinetic motion of the particle 
and fulfills the de Broglie relations [see eqn (\ref{eq-totsolu}) of \ref{App-A}], 
$k_d$ and its reciprocal, $\lam_d=2\pi/k_d$,  are apparently identifiable as the {\it de Broglie wavevector} and {\it  wavelength}.  Accordingly $\psiR$, or effectively ${\widetilde\Psim}$,  represents a traveling {\it de Broglie phase wave}.

Since the particle formed is in stationary state which is in mechanical terms only entailed by a standing wave mechanism,  $\psiR$ must therefore be looping in between the box walls repeatedly in many ($J$) loops; this corresponds to a total oscillation displacement of the charge being much greater than its radiation displacement $u^j$ commented earlier.
In effect $J$ needs be so large that the initially generated  wave prevail while the charge 
has, after a time $\tau_d (=2\pi/\w_d)$, turned to travel to the left at velocity $-\vel$,  which being virtual to the presently actual source. The virtual charge  generates 
similarly Doppler-differentiated (electromagnetic) waves 
$\varphi^{\dagsup {\rm vir}}= C_1 \sin(k^{\dagsup}_{-} X- \w^{\dagsup}_{-} T - \a_0+\beta_r)$, 
$\varphi^{\ddagsup {\rm vir}}
= -C_1 \sin(k^{\ddagsup}_{-} X+ \w^{\ddagsup}_{-} T + \a_0 +\beta_r)$. $\beta_r= k_d 2L  
+ (\w+\w_d) (\D T-\D T)=k_d 2L $ being a relative phase with respect to $\psiR$. 
$\varphi^{\dagsup {\rm vir}}$ 
and $\varphi^{\ddagsup {\rm vir}}$  superpose similarly to a beat wave, traveling to the left (dotted curve in Figure \ref{fig-dBwav}b): 
$$\displaylines{ \refstepcounter{equation}\label{eq-Eqvirt}
\qquad
\psiR^{{\rm vir}} ={\widetilde\Phim}^{{\rm vir}} {\widetilde\Psim}^{{\rm vir}}, \quad
{\widetilde\Phim}^{{\rm vir}}
=2C_1\cos[(K + (\vel/c) k_d) X  + (\vel/c)\w T], \  \hfill
\cr
\qquad {\widetilde\Psim}^{{\rm vir}}=- 
\sin[ k_d X + \w T+\a_0+\beta_r].   \hfill 
(\ref{eq-Eqvirt})}$$
$\psiR^{{\rm vir}} $, hence also ${\widetilde\Psim}^{{\rm vir}}$, has a phase velocity $-W$. 

For  $J>1 $ or practically  $J>>1$, the two opposite traveling beat waves $\psiR$ and $\psiR^{{\rm vir}}$ will meet at any time and location, and superpose  as $\psi =\psiR+\psiL$. 
For the case $\vel/c <<1$, we can ignore the source motion effect on  ${\widetilde\Phim} $ and ${\widetilde\Phim}^{{\rm vir}}$, and put   
${\widetilde\Phim} \simeq {\widetilde\Phim}^{{\rm vir}}\simeq  2C_1 \cos K X
$. 
Suppose the wave is of a mode fulfilling  $k_d 2L=n2\pi$, accordingly   $\w_d =k_d \vel =\frac{\hbar k_d ^2}{m} =\frac{\hbar (n2\pi/2L) ^2}{m}  $, with $n$ integers.
Or,  at the limit $(\vel/c)^2 \rightarrow 0$, 
$$\displaylines{
\refstepcounter{equation}\label{eq-kd1}
\label{eq-wd1}
\qquad 
K_d = n\pi/L,   \quad 
   \W_d 
= \hbar n^2 \pi^2 /(M L^2), \quad n=1,2,\ldots.  \hfill(\ref{eq-wd1}) 
}$$ 
Then, using again the trigonometric identity we obtain $\psi$ is a standing beat wave 
$
\psi(X,T)=\psiR+\psiR^{{\rm vir}}
= 4C_1\Phim \cos( k_d X +\alpha_0) \sin(- \w T)$,  with 
$\Phim /4=2 {\widetilde\Phim}/4 =2 {\widetilde\Phim}^{{\rm vir}}/4=\cos(K X).
$
At all other modes  the two waves annul, quite clearly for $J>>1$. 
Putting $\a_0=-\frac{\pi}{2}$ so that at the massive walls $\psi(0,T)=\psi(L,T)=0$ as is a mechanical requirement, 
 we have:   
$$\displaylines{\refstepcounter{equation}\label{eq-psi1}\label{eq-psi0}\label{eq-Eq8}
\qquad 
\psi=C\Phim \sin( k_d X ) \sin(- \w T).
 \hfill (\ref{eq-psi0})
}$$
Where  $C=4C_1/\sqrt{L}$ is a constant given after the normalization  $\int_0^L \psi^2 d X=1$. 

\subparagraph{\ref{Sec-V0.1}.4 The complex wave function}
Our operations hereafter will prominently involve 
 transformations between $\psi $ and  its derivatives such as $\frac{\pd \psi }{\pd T}$, $\frac{\pd^2 \psi }{\pd T^2}$, and $\frac{\pd^2 \psi }{\pd X^2}$,   
that are mutually orthogonal. For example, 
$\frac{\pd \psi }{\pd T}$
($\propto - \cos( \w T)$), 
is  orthogonal to $\psi$ $(\propto\sin( \w T))  $.
To ensure in the transformations 
the orthogonalities 
be formally preserved, it is necessary that we use the complex form of  $\psi$:
$$\displaylines{
\qquad
\psi =C e^{i[(K+k_d)X -\w T]} \quad {\rm or} 
                   \hfill (\ref{eq-psi1}{\rm a})
\cr
\qquad \psi  =C\Phim \Xim \Thm, \quad
\Phim= e^{iK X}, 
\quad 
        \Xim=e^{i  k_d X}, 
\quad 
       \Thm =e^{-i\w T}. 
 \hfill
}$$
For example,   $\frac{\pd \psi}{\pd T}=-i \psi $  turns, through the  prefactor $ -i$, the direction of  $\frac{\pd \psi}{\pd T}$ by $90$ degree clockwise 
from that of $\psi$,  and is indeed orthogonal to $\psi$, in either geometric terms or mathematical terms. The standing wave $\psi$ has in (\ref{eq-psi1}a) formally the form of a traveling plane wave; this is only as the immediate result of our use of the  complex form  to  reflect the orthogonalities. The real part of $\psi$ given in (\ref{eq-psi1}) describes the physical displacement which continues to be a standing wave.

\subparagraph{\ref{Sec-V0.1}.5 Total wave equation of particle }
For the differential equation (\ref{eq-CME1}) being linear, 
the linear combination of its solutions $\sum_j \varphi^j=\psi$, in the restricted manner so as to preserve the respective initial phases 
\footnote{
 Note that two waves generated by two independent sources, representing two particles, will not superpose in any coherent manner, since their initial phases are in general random with one another.
},  
is also a solution to it; this  $\psi$ is  just the total wave of the particle. 
We can therefore sum up the equations (\ref{eq-CME1}) over $j=\dagger, \ddagger, \dagger^{{\rm vir}}, \ddagger^{{\rm vir}}$ similarly, and obtain a classical wave equation for  the total wave of the particle: 
$$\displaylines{
 \refstepcounter{equation}\label{eq-CME1pp}\label{eq-eqmt1}
\qquad 
c^2 \frac{\partial^2  \psi(X,T)}{ \partial X^2}  
+ \frac{ V(X,T)}{m}\frac{\partial^2  \psi(X,T)}{ \partial X^2} =\frac{ \partial ^2 \psi(X,T)}{\partial T^2}.  
                    \hfill
                               (\ref{eq-CME1pp})
}$$
It can be verified (\ref{App-A}) that the explicit-superposition resultant solution for $\psi$ in (\ref{eq-psi1}a) substituted in  (\ref{eq-CME1pp}) gives correctly the established relativistic energy--momentum relation; this separately verifies that $\psi$ of (\ref{eq-psi1}a) is a solution to (\ref{eq-CME1pp}).

\section{Transformation to Schr\"odinger equation
}\label{Sec-V0}

We shall in the following separate out from (\ref{eq-CME1pp}) 
a component equation governing directly the particle's  kinetic motion,  through back-substitution of the  $\psi$  as obtained from explicit superposition given in (\ref{eq-psi1}a), and
combining with appropriate transformations and
simplifications, which  we will justify  where in question, for the classic-velocity limit. 
First, the $K$- and $K_d$- processes are explicitly separable  in $\frac{\pd  \psi }{\pd T} (\propto \w)$ and not in $\frac{\pd ^2 \psi }{\pd T^2} (\propto \w^2)$  (see \ref{App-A.1}). So, aimed to  
 separate the two processes, it is compelling that we make, 
for the known function of $\psi$ in  (\ref{eq-psi1}a),  
the transformation  
$$\displaylines{ 
\refstepcounter{equation}\label{eq-dpsidT}
\qquad
\frac{\pd ^2 \psi }{\pd T^2}= -i \g \W \frac{\pd  \psi }{\pd T}.
\hfill (\ref{eq-dpsidT})
}$$ 
Secondly, we here require $V(X,T)$, if not being constant, is slow varying only, such that everywhere in $L$ the particle $\psi$ (in quantized state) maintains the same eigen energy level.
Therefore for the $V$-term  we can replace the involved differential function
by its computed value: 
$\frac{\pd^2 \psi}{\pd X^2 }=(\frac{\pd^2 \Phim }{\pd X^2}) \Xim \Thm+ \Phim \frac{\pd^2 \Xim}{\pd X^2} \Thm
=-K^2 \g^2 \psi $, 
where for the final result we used the identity relation $(1+\g^2 \frac{\vel^2}{c^2})=\g^2$, and moreover also dropped  the cross-term between  
$ \frac{\pd \Phim}{\pd X}$  and $  \frac{\pd \Xim}{\pd X} $  that are mutually orthogonal (see \ref{App-cross}). 
Thus 
$ 
\frac{V}{m}   \frac{\pd^2 \psi}{\pd X^2 } 
=\frac{V }{\hbar}(-\frac{K^2 \g^2c^2}{c^2})\psi
=-\frac{V\w}{\hbar}\psi. $
Now substituting into (\ref{eq-CME1pp}) with the last equation  for the $V$-term,  with (\ref{eq-dpsidT}) for $\frac{\pd^2 \psi}{\pd T^2}$, and 
(\ref{eq-mM}) for $c^2$,    
  multiplying the resulting equation by
$-\frac{\hbar}{\w}$, we have:
$$\displaylines{
 \refstepcounter{equation}\label{eq-CME1p2}  
\qquad 
-\frac{\hbar^2}{m} \frac{\partial^2  \psi(X,T)}{ \partial X^2}  
+ V(X,T)\psi =i\hbar \frac{ \partial  \psi(X,T)}{\partial T}. 
 \hfill(\ref{eq-CME1p2})   
}$$

Thirdly, prepared with the above we shall proceed to separate the $(K,\W)$- and $(K_d,\W_d)$- processes by means of 
 expansion of the $\g$ factor, contained in (\ref{eq-CME1p2}) explicitly and inexplicably. 
It is easy to see that  (\ref{eq-CME1p2})  is the differential form of an energy equation;  
and the computed value of each its terms  
  contains  a factor $\g$.   
As analyzed in \ref{App-A}, 
in the   expansion of $\g$,  
the order $\vel^2/c^2$ -term  
corresponds to
the particle's  kinetic energy 
which is 
a {\it classical} quantity, not a "hyperclassic-velocity" or "relativistic" correction. This term must therefore  be maintained when taking the classic-velocity limit $\vel^2/c^2\rightarrow 0$. 
Prior to that we must firstly expose the $\g$ factors of all terms at the same level,  by making the following operations; 
we first compute the terms of (\ref{eq-CME1p2}) by substituting  into them with (\ref{eq-psi1}a) for $\psi$.  
The two differential terms are  
$\frac{1}{m} \frac{\pd^2 \psi}{\pd X^2} 
=-\frac{\g^2 K^2\psi}{\g M}
$ 
and 
$\frac{\pd \psi}{\pd T}=-i \g \W \psi 
 $;
expanding the  $\g$ factor, these write:  
 $$\displaylines{\refstepcounter{equation}\label{eq-exact1}
\qquad 
\frac{1}{m} \frac{\pd^2 \psi}{\pd X^2} 
=-\frac{K^2\psi }{M}        -\lf(\frac{K_d^2}{2M}+\ldots\rt) \psi, 
\quad
  \frac{\pd \psi}{\pd T}
= -i (\W+\frac{1}{2}\W_d+\ldots)\psi. \hfill
}$$
In the two expansions the leading   terms,  $\propto K^2/M$ and $\propto\W$, represent each 
the particle's total rest energy.  The remaining terms in series  give each the particle's relativistic kinetic energy, with the leading terms 
$\frac{K_d^2}{2M}  $ and $\frac{1}{2}\W_d$ being values at the classic-velocity limit.  

Fourth, 
we now take the classic-velocity limit  $\vel^2/c^2  \rightarrow 0 $ of the above two differential functions, 
to remove the "relativistic" corrections only while maintain the classical quantities dependent on $\vel^2/c^2$.
This  as just analyzed amounts to take the classic-velocity limit of the kinetic terms in the expansions:  
$$\displaylines{
\qquad
\lim_{\vel^2/c^2 \rightarrow 0}\frac{1}{m} \frac{\pd^2 \psi}{\pd X^2} 
= -\frac{K^2 }{M} \lim_{\vel^2/c^2 \rightarrow 0}\psi
      -
\lim_{\vel^2/c^2 \rightarrow 0} \lf(\frac{K_d^2}{2M}+\ldots\rt) \psi
=-\frac{K^2\Psim }{M}        - \frac{K_d^2\Psim}{2M} ,   \hfill
\cr
\qquad
 \lim_{\vel^2/c^2 \rightarrow 0}
 \frac{\pd \psi}{\pd T}
= 
-i \W\lim_{\vel^2/c^2 \rightarrow 0}\psi 
- \lim_{\vel^2/c^2 \rightarrow 0}    i( \\Omegavel +\ldots)\psi
= -i \W\Psim -i \Omegavel \Psim.  
\hfill
(\ref{eq-exact1})
}$$
Where $\Omegavel \equiv \frac{1}{2} \W_d$; and
 $$
\displaylines{\refstepcounter{equation}\label{eq-conv0}
\qquad
\Psim =\lim_{\vel^2/c^2 \rightarrow 0}\psi 
=C e^{i K_d X} e^{-i  \Omegavel T}.
\label{eq-aprox1}
  \hfill (\ref{eq-conv0})
}
$$
For the final result of (\ref{eq-conv0}) we used the limit relations 
$\lim_{\vel^2/c^2 \rightarrow 0} Re[\Phim] = 1$,  and 
  $
\lim_{\vel^2/c^2 \rightarrow 0} \Thm(T)
= $ 
$\lim_{\vel^2/c^2 \rightarrow 0} $ $
e^{-i (\W+ \Omegavel)T  } $ 
$
=  e^{-i \Omegavel T} $; 
these reflect that  at the  scale of the de Broglie wave parameters $K_d$ and $\Omegavel $, 
the oscillations associated with the $K$ and $\W$ variables are no different from constants, as we can  directly see in Figure \ref{fig-dBwav}.
Differentiating 
(\ref{eq-conv0}) directly  we have
$
 \frac{1}{M}\frac{\pd^2 \Psim}{\pd X^2} =-\frac{K_d^2}{M} \Psim$, 
$
  \frac{\pd \Psim}{\pd T}
  = -i  \Omegavel \Psim $. 
Substituting the last two relations for the  corresponding computed values on the right-hand sides of the two relations of (\ref{eq-exact1}), we obtain the corresponding  results in differential function forms    at the classic-velocity limit. In turn with the resultants in (\ref{eq-CME1p2}) we obtain the corresponding differential equation 
$$\displaylines{\refstepcounter{equation}\label{eq-schtot}
\qquad 
\frac{\hbar^2 K^2}{ M} \Psim - \frac{\hbar^2}{2M}\frac{\pd^2 \Psim }{\pd X^2}
+ V\Psim=-\i^2 \hbar \W \Psim +  i\hbar  \frac{\pd \Psim}{\pd T}.
\hfill  (\ref{eq-schtot}) 
}$$
Subtracting the  equation of total rest energy and total rest momentum, $\frac{\hbar^2 K^2}{M}
=\hbar \W$ multiplied by $\Psim $ on both sides, (\ref{eq-schtot}) thus reduces to an equation directly describing the kinetic motion of the particle at the classic-velocity limit:
$$\displaylines{ \refstepcounter{equation}\label{eq-sch}
\qquad 
 - \frac{\hbar^2}{2M}\frac{\pd^2 \Psim }{\pd X^2}
+ V\Psim=  i\hbar  \frac{\pd \Psim}{\pd T}. 
\hfill (\ref{eq-sch})
}$$
This we see is equivalent to the Schr\"odinger equation for an identical system, a single particle  bound in a potential $V$, with $V=$ (or $\approx$) $V_0$ for the above derivation. 
We also readily recognize  that, 
(\ref{eq-sch}) if solved with the  standard quantum-mechanical formalism give solutions  for eigen state wave function $\Psim$ and variables like 
 $K_d$, $\Omegavel $, and $\Eng_\vel$ 
that are equivalent to our classical-mechanical results given by  (\ref{eq-conv0}),    (\ref{eq-kd1}), and  (\App\ref{eq-Ev1}).

\section{Schr\"odinger equation in variant potential. Plane-wave approach}
For $V(X,T)$ being arbitrarily variant,  the component total waves $\varphi^j$, likewise the $\Psim$, are no longer each plane waves across the  box $L$. However the corresponding partial waves  as generated in any infinitesimal region say $(X_i, X_i+\D X)$ where  $V_i = V(X_i,T)$ is effectively constant, continue to be so. So, to utilize the earlier plane wave solutions as an effective means in the above sense,  we divide $L$ in $N$ small divisions, and express  $V(X,T)$, assuming well behaved, with the stepwise potential: $V_i=f_i V(X_i,T), i=1, 2, \ldots N$, where $f_i=1$ for $X\in (X_i, X_i+\D X)$ and  is zero for $X$ elsewhere. $V_i$  has a notable  property: 
$$\displaylines{\refstepcounter{equation}\label{eq-Vs}
\qquad
\sum {}_{i=0}^{N} V_i=V(X,T) 
\hfill (\ref{eq-Vs})
}$$
Within an $i$th division  we thus have the plane wave solutions $\varphi^j_i $ as of  (\ref{eq-Eq3}),  
 $\psi_i$ as of (\ref{eq-Eq8}),   
$\Psim_i =C_i \sin(K_{di} X) e^{-i\frac{1}{2}\W_{di}T} $ and  
$\Eng_{v_i }-V_i=\frac{(\hbar K_{d_i})^2}{2M} $ as of (\ref{eq-conv0}) and (\App\ref{eq-EPvel}) or (\App\ref{eq-Ev1}), 
but with $K$, $\W$, $K_{d}$, and $\W_{d}$ etc, replaced by $K_i$, $\W_i$   $K_{di}$, and $\W_{di}$ for the $i$th division here.
The total wave equation for $i$th division writes as of (\ref{eq-psi1}{\rm a}), and after  a similar reduction, separates out an equation as of (\ref{eq-sch}) for each $\Psim_i$. Multiplying on both sides with a weight coefficient $A_{i}$, summing the equations for all $i$ we have:
$$\displaylines{\qquad
 -\frac{\hbar^2}{2M} \frac{\partial^2 \sum_i A_{i}\Psim_i(X,T)}{ \partial X^2}  +\sum_i V_i \sum_i A_{i}\Psim_i (X,T)=i\hbar \frac{ \partial  \sum_i A_{i}\Psim_i(X,T)}{\partial T} \hfill({\rm c}).
}$$
Where we have put  $\sum_i A_iV_i \Psim_i  
=\sum_i V_i \sum_i A_i\Psim_i $ on the grounds that $V_i $ and $\Psim_{i'} $ are uncorrelated for $i\ne i'$. $\sum_i V_i $ sums according to (\ref{eq-Vs}). Now $\Psim_i$ is generated when the source charge is in $(X_i, X_i+\D X)$; its dynamical variables $\Psim_i$,  $K_{di}$, $\Eng_{vi}$, etc. as reflected  above are all characterized by the $V_i$ in it.
 On the other hand,  the resulting $\Psim_i$ as a mechanical wave process, as with the forces on the  medium,  is nonlocal for the uniform continuum of vacuum within $L$, which  presents  no "walls"  or "scatterers" to reflect or deflect the wave. In other words, this medium will transmit the force acting  at one location to all others, and thereby the deformation of the medium at one location to all others, the latter being through wave propagation. So $\Psim_i$ must be propagated uniformly across $L$ and reflected only at  the box walls at $0, L$. As a result,
 at any location in the box we will find the series of  $\{\Psim_i\}$ generated  from all divisions; the total wave is  thus: 
$$\displaylines{\refstepcounter{equation}\label{eq-PsimVXT}
\qquad
\Psim(X,T)=\sum_iA_i \Psim_i (X,T). 
\hfill                                  (\ref{eq-PsimVXT})
}$$
 With (\ref{eq-PsimVXT}) for $\sum A_i\Psim_i$ and (\ref{eq-Vs}) for $\sum V_i$ in ({\rm c}),
we get precisely a  Schr\"odinger equation as of (\ref{eq-sch}) in terms of $\Psim$ and $V(X,T)$, here for  $V(X,T)$ arbitrarily variant  in $(0,L)$ and well behaved.
We may in principle obtain the eigen solutions from summing the respective plane-wave results, e.g.  
for energy
   $\sum_i (A_i \Psim_i )^2 [(\Eng_{vni }-V_i)-\frac{(\hbar K_{dni})^2}{2M} ]=0
$, but we can continue to, efficiently, solve 
(\ref{eq-sch}) using the established quantum-mechanical formalism. 

\section{Schr\"odinger equation in three dimensions}

In a three-dimensional space, the point-like source charge of a particle, at a given location $\Rb(T)$, generates radial waves with a  donut-ring stereo contour. However in view that the total wave energy and also the resulting vacuum polarization are independent of the radial distance  $\lb=\Rb'-\Rb $ from  $\Rb$, the waves can be represented as a plane wave on an apparent linear chain along $\lb$ (see a full account given in [\citeRef1 b]).  Furthermore,  
          %
at any given $\Rb$ and $\vb(\Rb,T)$, the particle  develops a de Broglie wave only along the direction of its instantaneous motion, $\vb$, that is along the tangent  $\lb(\Rb)$;   at the limit $\lb \rightarrow 0$,   $\lim_{
 \Rb' \rightarrow\Rb
} \lb = \Rb $.  As a result, the problem reduces to that for a local linear chain and plane wave. About any  $\Rb$, the inference  of  Schr\"odinger equation in terms of the plane wave approach in the preceding sections holds provided we replace  $X$ with the present vector distance $\Rb$, and $\frac{\partial^2}{\partial X^2}$ with $\nabla^2 $: 
$$\displaylines{\refstepcounter{equation}\label{eq-CME4pp}
\qquad
 -\frac{\hbar^2}{2M}\nabla^2 \Psim(\Rb,T)  + V(\Rb,T) \Psim (\Rb,T)=i\hbar \frac{ \partial  \Psim(\Rb,T)}{\partial T}.    \hfill(\ref{eq-CME4pp})
}$$ 
So we remain only the need to transform the independent, coordinate variable $X$ to $\Rb$ and the derivatives  with respect to $X$ to $\Rb$, entirely as we used to do in quantum mechanics.
 In terms of Cartesian coordinates e.g. we have $\Rb=X \hat{X}+Y\hat{Y}+Z\hat{Z}$, 
$\nabla^2 = \frac{\partial^2}{\partial X^2}  + \frac{\partial^2}{\partial Y^2}+\frac{\partial^2}{\partial Z^2}$.


Some of the improvements in this paper have benefited from discussions at the Varna QTS-4; a more detailed acknowledgement is given in [\citeRef1 a].


\section*{Appendix}
\setcounter{equation}{0}
\setcounter{section}{0}

\begin{appendix} 
\section{Relativistic energy-moment relation from the total wave equation and solution
}  \label{App-A}

\subparagraph{The quadratic form} \label{App-A.1}
  Substituting $\psi$ of (\ref{eq-psi1}a) and its derivatives in  (\ref{eq-eqmt1}), dropping the cross-term $\frac{\pd \Phim}{\pd X}\frac{\pd \Xim}{\pd X}$ as justified in  \ref{App-cross},
and multiplying $\frac{\hbar^2}{i^2}$ through,   we have
$$\displaylines{
 \refstepcounter{equation}\label{eq-enga}
\qquad 
\hbar^2c^2 (1+V/mc^2) K^2 [1+\g^2(\vel^2/c^2)]
=\hbar^2\W^2 \g^2.
\hfill (\App\ref{eq-enga})
}$$
Put   $\eng=\hbar \g \W $ and  
 $p_\vel=  m \vel $, these being the total relativistic  energy and the   linear momentum of the particle. 
 Substituting into  (\App\ref{eq-enga}) with these together with the identity relation $\hbar K=Mc$, accordingly 
$M^2c^2   (1+\g^2 \frac{\vel^2}{c^2})=
M^2c^2  +p^2_\vel $, we have 
$$\displaylines{
\refstepcounter{equation}\label{eq-engb}
\qquad
 (M^2 c^4 +p^2_\vel c^2) (1+V/mc^2) =\eng^2.
\hfill
(\App\ref{eq-engb}) 
}$$
(\App\ref{eq-engb}) or equivalently (\App\ref{eq-enga}) is seen to be  the well-established relativistic energy-momentum relation in the more general case of a finite  $V$---that this is fulfilled  by $\psi$ gives an independent verification that $\psi$ is a  solution to (\ref{eq-CME1pp}).

\subparagraph{The linear form}  

We want now to convert the quadratic  equation (\App\ref{eq-engb}) to linear form; the analysis to be given  
underlines the reason why we transformed the $\frac{\pd^2 \psi}{\pd T^2}$ in  (\ref{eq-CME1pp}) to 
$\frac{\pd \psi}{\pd T}$ as in (\ref{eq-CME1p2}). 
If taking the square root of (\App\ref{eq-engb}) on each side  we would have    
$ \sqrt{ (M^2 c^4 +p^2_\vel c^2)}
\sqrt{(1+V/mc^2)} = \pm \eng
$.  
  Despite its right-hand side ($\eng$) is the same as that of the (\App\ref{eq-eng1}) below, 
   however its left-hand side  is a nonlinear superposition  between the $K$- and $K_d$- terms, and consequently 
between the force terms $F_\re$- and the $F_\med$-  depending in general  of $K$- and $K_d$. This is   undesirable for our purpose of separating the two  processes. 
So instead, we choose to 
divide first by $ \hbar \W \g$ the equation (\App\ref{eq-enga}). After rewriting  as 
$\hbar^2 K^2 c^2 \g^2 + \frac{\hbar^2K^2c^2\g^2 V}{Mc^2} =\hbar^2 \W^2 \g^2$, 
where   $ \frac{\hbar^2K^2c^2\g^2 V}{Mc^2 \hbar \W \g}=V$, 
this gives 
$$
\refstepcounter{equation}\label{eq-eng1a}
\hbar  K c\g +V =\hbar \W \g. 
\eqno(\App\ref{eq-eng1a})
$$
We next expand the $\g$ of the two remaining  terms   of (\App\ref{eq-eng1a}),  
$\g=1/\sqrt{1-\vel^2/c^2}
=1+\frac{1}{2}\frac{\vel^2}{c^2}+\frac{3}{8}\frac{\vel^4}{c^4} +\ldots$, reorganize combined with use of the identity relations  $\hbar K =Mc$ and $K(\vel/c)=K_d$, $\frac{1}{2}\W(\vel/c)^2= \Omegavel $, 
and put 
$$\displaylines{
 \refstepcounter{equation}\label{eq-totsolu}
\hfill P_\vel=\hbar K_d, \quad \Eng_\vel=\hbar \Omegavel.  
\hfill(\App\ref{eq-totsolu}) 
}$$  
With the above, (\App\ref{eq-eng1a}) writes 
$$\displaylines{ 
\refstepcounter{equation}\label{eq-eng1}
\hfill
Mc^2 + \frac{P_\vel^2}{2M} \lf(1+ \frac{3}{4}\frac{\vel^2}{c^2} +\ldots\rt)
+V
=\Eng +\Eng_\vel  \lf(1+ \frac{3}{4}\frac{\vel^2}{c^2}+\ldots  \rt). 
\hfill (\App\ref{eq-eng1})
    }$$ 

\subparagraph{The classic-velocity limit}  

Taking the classic-velocity limits of the two kinetic terms, that is  the $P_\vel$- and $\Eng_\vel$- terms  multiplied with a series each  in
(\App\ref{eq-eng1}), we have 
$$\displaylines{
 \refstepcounter{equation}\label{eq-WPp}
\qquad
 M c^2 +\frac{P^2_\vel}{2M} +V =\Eng +\Eng_\vel. \hfill
(\App\ref{eq-WPp}) 
}$$
Either side of it is a total rest energy plus a classic-velocity limit of the  mechanical  energy in respect to the kinetic motion of the particle. That is, (\App\ref{eq-WPp}) is a total energy equation at the classic-velocity limit.

\subparagraph{Newtonian energy-momentum relation}  
Subtracting  the total rest energy-momentum relation $(Mc)c=\Eng$  from (\App\ref{eq-WPp}) therefore gives 
$$\displaylines{
 \refstepcounter{equation}\label{eq-EP}\label{eq-EPvel}
\qquad 
\Eng_\vel=\frac{P_\vel^2}{2m}+V,
  \hfill(\App\ref{eq-EPvel})
}$$ 
which the Newtonian energy-momentum relation.   

\subparagraph{Standing-wave solutions for dynamical variables  of particle }
With the standing-wave result  (\ref{eq-kd1}) for $K_d$ 
in (\App\ref{eq-EPvel}),  suffixing it by $n$ as $K_{dn}$ for distinction below,  we have the standing-wave solution for mechanical energy
$$\displaylines{
\refstepcounter{equation}\label{eq-Ev1}
\qquad
\Eng_{\vel n}=(\hbar K_{dn})^2/(2M) +V
= n^2 h/(8ML)+V, \qquad n=1,2,3,\ldots 
\hfill(\App\ref{eq-Ev1})
}$$
The solutions $ K_{dn}$ and $\Eng_{\vel n} $ given in (\ref{eq-kd1}) and (\App\ref{eq-Ev1}) in terms of classical mechanics we see are all quantized for $L \sim$ $\Lam_{dn}$. As we have seen through the treatment, the quantization of dynamical variables of the particle are the  direct result of standing wave solutions.

\section{Orthogonality between the cross terms }
\label{App-cross}

The processes described by $\Psim(X,T)$ and $\psi(X,T)$ are both of statistical nature, so must be the dynamical variables weighted by them and their derivatives as are in the two wave equations  (\ref{eq-sch})  and (\ref{eq-CME1pp}). 
Thus, it ought to be the "thermal average" here indicated by $<\ >$ of each of the  terms that yield the corresponding relations of the dynamical variables of particle. 
This 
{\it rule} we see corresponds just to the Ehrenfest's theorem 
for equation (\ref{eq-sch}) which corresponds to the Schr\"odinger equation,
and $<\ >$ corresponds to 
the expectation values.\footnote{
This will be equivalent to the expectation value in the quantum mechanical formalism, indicated here by $< \ >_{{\rm qm}}$;  e.g. consider $\Psim(X,T)$, 
for the variable $K_d^2$ this is  
$<\frac{\pd^2 \Psim}{ \pd X^2} >_{{\rm qm}}
=\int \Psim^* \frac{\pd^2 \Psim}{ \pd X^2} d X
= |\Psim|^2 i^2 K_d^2 \int_0^L d X =-C^2 L$. 
Here the time-dependent phase factor is a complex exponential $e^{-i \W T}$ and thus annul to 1 always in the QM the definition of probability density $|\Psim|^2$. So a time average may be seen as  effectively done in this way and needs not be  explicitly involved.  
}
Similarly, with an application of the above {\it rule} to equation   (\ref{eq-CME1pp}), we have the corresponding equation for the  respective thermal averages  
$$\displaylines{
 \refstepcounter{equation}\label{eq-eqmt1Ap}
\qquad 
\lf<c^2 \frac{\partial^2  \psi(X,T)}{ \partial X^2}  \rt>
+ \lf<\frac{ V(X,T)}{m}\frac{\partial^2  \psi(X,T)}{ \partial X^2}\rt> 
=\lf<\frac{ \partial ^2 \psi(X,T)}{\partial T^2}\rt>.  
                    \hfill
                               (\App\ref{eq-eqmt1Ap})
}$$
Here,  $<A>=\int_{L_1} \int_{T_1} A d X dT $, with $L_1$ and $T_1$ being a macroscopic length and time. 
With this final solution as our only interest, we can simplify the computation of the $\frac{\pd^2 \psi}{ \pd X^2}   $ term as follows.
We first expand this term, after a division  of $\psi$ as  
 follows from dividing out $\psi$ on both sides of  (\App\ref{eq-eqmt1Ap}), and get
$$\displaylines{
\refstepcounter{equation}\label{eq-divap1}
  \qquad
 \frac{ 1}{\psi} \frac{\pd^2 \psi}{ \pd X^2}   
= \frac{1}{\psi} 
[(\frac{\pd^2 \Phim}{ \pd X^2}) \Xim
        +\Phim\frac{\pd^2 \Xim}{ \pd X^2} 
        +2 (\frac{\pd \Phim}{ \pd X})(\frac{\pd \Xim}{ \pd X})] \Thm \hfill
\cr
\qquad 
=\frac{1}{\psi}
 [i^2K^2(T)  + i^2 K_d^2(T) +2i K(T) iK_d(T) ]\psi \hfill
\cr
\qquad 
=- [K^2(T)  +  K_d^2(T) + 2K(T) K_d(T) ]
\hfill (\App\ref{eq-divap1})
}$$
Taking the thermal average of  (\App\ref{eq-divap1}) 
  over $\Lam_d$ and $\Taum_d/2$ as suffices for $\psi$ being periodic, we have 
$$\displaylines{\refstepcounter{equation}\label{eq-themav}
\qquad
\frac{1}{C^2}\lf< \frac{ 1}{\psi} 
\frac{\pd^2 \psi}{ \pd X^2}    \rt>
=\frac{1}{C^2\Lam_d(\Taum_d/2)} 
\int_0^{\Lam_d} 
\int_{0}^{\Taum_d/2 }
\frac{ |\psi|^2}{\psi} 
\frac{\pd^2 \psi}{ \pd X^2} d X d T  \hfill
\cr
\qquad =-
\frac{1}{\Lam_d(\Taum_d/2)}  [ X]_0^{\Lam_d} \int_{0}^{\Taum_d/2 }  
            [K^2(T)  +  K_d^2(T) + 2K(T) K_d(T) ]
       d T \hfill
\cr
\qquad =-\frac{1 }{\frac{N\Taum}{2}} 
\lf\{  
            \{ [K^2  +  K_d + 2K K_d]T\}_0^{\Taum/2}
       +  
            \{ [(-K)^2  +  K_d^2 + 2(-K) K_d] T\} _{\Taum/2}^{\Taum}     
\rt\}\times \frac{N}{2}
 \hfill
\cr
\qquad=-\frac{1 }{\Taum} 
\lf\{  
            [K^2 (\frac{\Taum}{2}+\frac{\Taum}{2})  
+  K_d (\frac{\Taum}{2}+\frac{\Taum}{2}) 
+ 2 K K_d( \frac{\Taum}{2} - \frac{\Taum}{2}) ]       
\rt\}\hfill 
\cr
\qquad 
=-(K^2+K_d^2), \hfill (\App\ref{eq-themav})
}$$
where $|\psi|^2=C^2$, 
$\frac{\Taum_d/2}{\Taum}= \frac{N}{2}$. In one half of the time, i.e. over a duration $\Taum_d/2$,  the wave component $\Phim$ is traveling in $+X$-direction with $K>0$, and  in the other one half  time in the $-X$-direction with $K<0$. So, 
  the positive and negative cross-terms  between $\frac{\pd \Phim}{\pd X}$ and $\frac{\pd \Xim}{\pd X}$ annul on average and gives  zero net contribution to the thermal average,  as given by the final result of (\App\ref{eq-themav}). In other words,  the functions 
$\Phim$ and $\Xim$ are mutually orthogonal.  

\end{appendix}

\end{document}